\documentclass[9pt, aps, pra, twocolumn, superscriptaddress]{revtex4-2}

\usepackage[utf8]{inputenc}
\usepackage[T1]{fontenc}

\usepackage{graphicx}
\usepackage{physics}
\usepackage{amsmath}
\usepackage{multirow}
\usepackage{xcolor}
\usepackage[left=2cm,right=2cm,top=2cm,bottom=2cm]{geometry}
\usepackage{comment}

\usepackage{hyperref}

\begin{document}

\title{Quantum Link Prediction in Complex Networks}


\author{João P. Moutinho}
\altaffiliation{Corresponding authors: joao.p.moutinho@tecnico.ulisboa.pt}
\affiliation{Instituto Superior Técnico, Universidade de Lisboa, 1049-001 Lisboa, Portugal}
\affiliation{Instituto de Telecomunicações, 1049-001 Lisboa, Portugal}
\author{André Melo}
\affiliation{Kavli Institute of Nanoscience, Delft University of Technology, 2628 CD Delft, The Netherlands}
\author{Bruno Coutinho}
\affiliation{Instituto de Telecomunicações, 1049-001 Lisboa, Portugal}
\author{István A. Kovács}
\affiliation{Department of Physics and Astronomy, Northwestern University, Evanston, 60208 IL, USA}
\affiliation{Northwestern Institute on Complex Systems, Northwestern University, Evanston, 60208 IL, USA}
\affiliation{Central European University, 1051 Budapest, Hungary}
\author{Yasser Omar}
\affiliation{Instituto Superior Técnico, Universidade de Lisboa, 1049-001 Lisboa, Portugal}
\affiliation{Portuguese Quantum Institute, 1049-001 Lisboa, Portugal}
\affiliation{Centro de Física e Engenharia de Materiais Avançados (CeFEMA), Physics of Information and Quantum Technologies Group, 1049-001 Lisboa, Portugal}

\date{\today}

\begin{abstract}
\noindent Predicting new links in physical, biological, social, or technological networks has a significant scientific and societal impact. Path-based link prediction methods utilize explicit counting of even and odd-length paths between nodes to quantify a score function and infer new or unobserved links. Here, we propose a quantum algorithm for path-based link prediction, QLP, using a controlled continuous-time quantum walk to encode even and odd path-based prediction scores. Through classical simulations on a few real networks, we confirm that the quantum walk scoring function performs similarly to other path-based link predictors. In a brief complexity analysis we identify the potential of our approach in uncovering a quantum speedup for path-based link prediction.
\end{abstract}

\maketitle

\section{Introduction}

From genes and proteins that govern our cellular function, to our everyday use of the Internet, Nature and our lives are surrounded by interconnected systems \cite{barabasi2016network}. Network science aims to study these complex networks, and provides a powerful framework to understand their structure, function, dynamics, and growth. Studies in network science typically have a substantial computational component, borrowing tools from graph theory to extract relevant information about the underlying system. With the advent of quantum computation, a natural question to ask is which problems in network science can be explored with this new computing paradigm, and what benefits it can yield. This question can be interpreted in at least two different ways. First, there is a large body of work in quantum algorithms for graph theoretical problems, some examples being Refs. \cite{durr2006quantum, ambainis2006quantum, chakraborty2016spatial, chakraborty2017optimal}, which may have their own applications in network science problems. However, network science algorithms often look for specific patterns or organizing principles based on empirical observations from the real underlying systems, which may warrant the development of problem-specific quantum algorithms. This constitutes a novel research direction, different from the development of more general graph-theoretical algorithms. Previous connections have been made between quantum phenomena and complex networks, both by using quantum tools to study complex networks \cite{tsomokos2011quantum, sanchez2012quantum, faccin2013degree, mukai2020discrete} and by using complex network tools to study quantum systems \cite{faccin2014community}. Nevertheless, to our knowledge, potential quantum speedups for network science problems have not been addressed.

In this work we propose a quantum algorithm to the problem of link prediction in complex networks using Continuous-Time Quantum Walks (CTQW) \cite{farhi1998quantum, kempe2003quantum} inspired by popular path-based methods, and discuss potential quantum speedups over classical algorithms. The objective in link prediction is to identify unknown connections in a network \cite{Liben:2007,  wang2015link, albert2004conserved, getoor2005link, Lu:2011, zhou2021progresses}. For example, in social networks, we aim to predict which individuals will develop shared friendships, professional relations, exchange of goods and services or others \cite{Liben:2007, wang2015link}. In biological networks, the main focus is the issue of data incompleteness, which hinders our understanding of complex biological function. For example, in protein-protein interaction (PPI) networks link prediction methods have already proven to be a valuable tool in mapping out the large amount of missing data \cite{kovacs2019network, luck2020interactome}. While there are many approaches to the problem of link prediction \cite{zhou2021progresses}, such as using machine learning techniques \cite{al2006link, ghasemian2020stacking}, stochastic block models \cite{guimera2009missing} or studying global perturbations \cite{lu2015toward}, other methods focus on simple topological features like paths of different length between nodes quantified by powers of the adjacency matrix. Path-based methods are simple but remarkably popular, and have been shown to be competitive with other approaches in networks of various types \cite{zhou2021progresses, zhou2021experimental, muscoloni2022adaptive, muscoloni2021short}. In our work we show that quantum walks can be used as a natural encoding of path-based link prediction in the development of a quantum algorithm.

\begin{figure*}[t]
    \centering
    \includegraphics[width=0.7\textwidth]{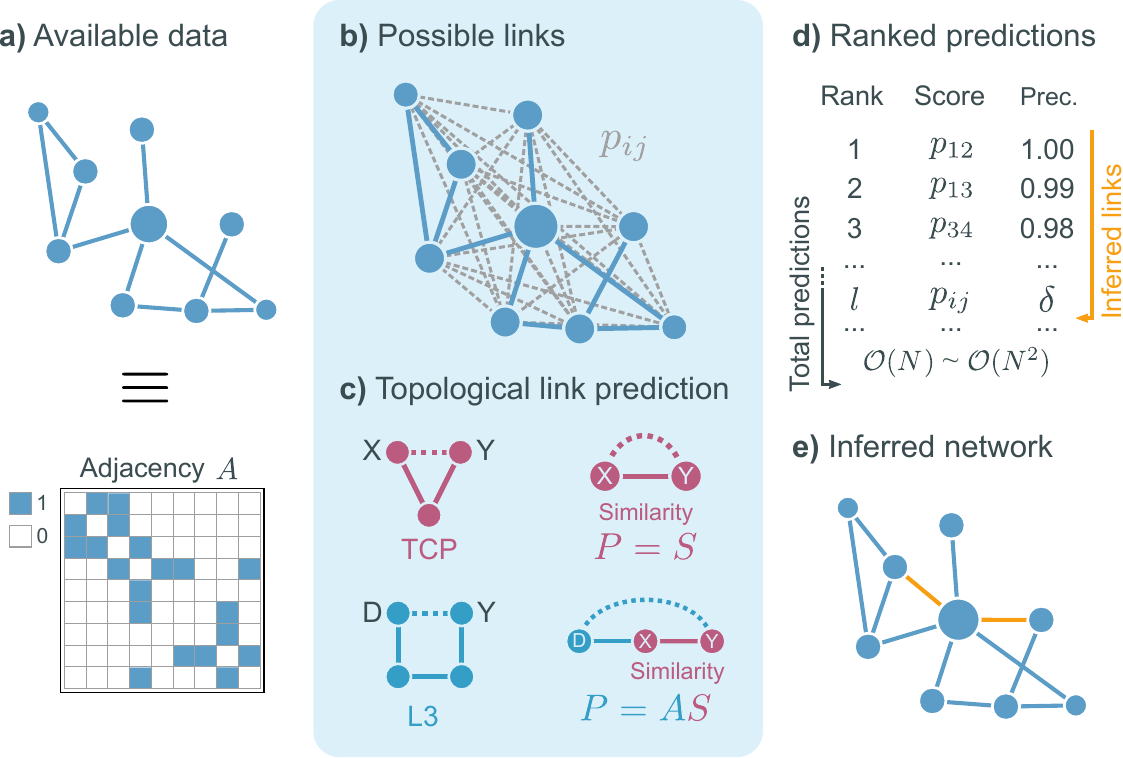}
    \caption{\textbf{Classical path-based link prediction.} {\textbf{a)} Link prediction methods take as input a complex network with a corresponding adjacency matrix $A$. \textbf{b)} Each method then associates a prediction value $p_{ij}$, or \textit{score}, to every pair of nodes \{i, j\}, such that a higher value $p_{ij}$ correlates to a higher probability of the link \{i, j\} appearing. \textbf{c)} This requires assumptions about the organizing principles of each network. Predictions based on the Triangle Closure Principle (TCP) rely on similarity between nodes, represented as a matrix $S$, a common assumption about connections in social networks. This can be quantified in the simplest case as $P\sim S\sim A^2$, counting paths of length 2 between pairs of nodes. As an alternative, proteins often connect to others that are similar to their neighbours, but not necessarily similar to themselves, quantified for example as $P\sim AS\sim A^3$, counting paths of length 3 between nodes \cite{kovacs2019network}. \textbf{d)} Most classical link prediction methods output all non-zero entries of matrix $P$, organized in a ranked list of scores from highest to lowest, where the relevant top $l$ predictions are those where the precision is above a user-determined threshold $\delta$. \textbf{e)} The relevant predictions are considered as new inferred links, represented in yellow, while the rest are discarded.}}
\label{fig:1}
\end{figure*}

\section{Classical Path-Based Link Prediction}
\label{sec:clp}

We start with a brief review of path-based link prediction. Link prediction methods take as input a graph $G(\mathcal{V},\,\mathcal{E})$, where $\mathcal{V}$ is the set of nodes with size $N=|\mathcal{V}|$ and $\mathcal{E}$ is the set of undirected links, and output a matrix of predictions $P\in\mathbf{R}^{N\times N}$ where each entry $p_{ij}$ is a score value quantifying the likelihood of a link existing between nodes $i$ and $j$ (see Figure \ref{fig:1}). Each method computes $P$ differently, depending on the assumptions made about the network and its emergent topological features. Most path-based methods are based on the Triadic Closure Principle (TCP), assuming that two nodes are more likely to connect the more similar they are \cite{Lu:2011, kovacs2019network}. Given a matrix $S\in\mathbf{R}^{N\times N}$ quantifying similarity between any two nodes, predictions based on TCP assume that
\begin{equation}
P=S.
\end{equation}
Similarity is often quantified based on the number of shared connections, i.e., paths of length two between two nodes, which can be computed as $S=A^2$. A possible generalization is to consider a linear combination of even powers of $A$. It has been shown that, despite its dominant use in biological networks, the TCP approach is not valid for most protein pairs \cite{kovacs2019network}. Instead, in \cite{kovacs2019network}, a link prediction method (L3) is proposed without the assumption that node similarity correlates with direct connectivity. L3 is based on the assumption that a potential new link $(i,\,j)$ relies on $i$ being similar to the existing neighbours of $j$. In matrix form, predictions based on the L3 paradigm may be computed by extending the similarity matrix $S$ one step over the adjacency matrix, 
\begin{equation}
P=AS,
\end{equation}
as illustrated in Figure \ref{fig:1} c). Considering the simple case of $S=A^2$, the authors in \cite{kovacs2019network} define $P$ based on $AS=A^3$, with an added degree normalization. Their results show the L3 method significantly outperforms other TCP-based methods in the prediction of protein-protein interactions. At the same time, the LO method was proposed in \cite{pech2019link}, which represents $P$ as a linear combination of odd powers of $A$, also showing significant improvements over TCP-based methods. Other follow-up studies proposed different L3-based methods \cite{Cannistraci:2018}, and further studied the application of L3- and TCP-based methods \cite{kitsak2020latent, zhou2021experimental}, concluding that L3-based methods perform well in various network categories.

Our quantum approach takes inspiration from both these paradigms, utilizing even (TCP) and odd (L3-like) powers of $A$. One of the main reasons why link prediction may prove suitable to be tackled with a quantum computer is the realisation that in practice we are not interested in knowing the scores of all pairs of nodes, but we simply wish to know which ones have the highest score up to a certain cut-off threshold, as illustrated in Figure \ref{fig:1} d) and e). By encoding the prediction scores in the amplitudes of a quantum superposition and performing quantum measurements on the system, the predictions with the highest score will be naturally sampled with higher probability, which can potentially be advantageous compared to the classical case of explicitly computing all scores as long as the quantum superposition can be efficiently prepared. We proceed now in Section \ref{sec:quantum} with the description of the quantum method, and discuss in Section \ref{sec:results} the expected resource complexity and show example comparisons with classical path-based methods.

\begin{figure*}[t]
\centering
\includegraphics[width=0.8\textwidth]{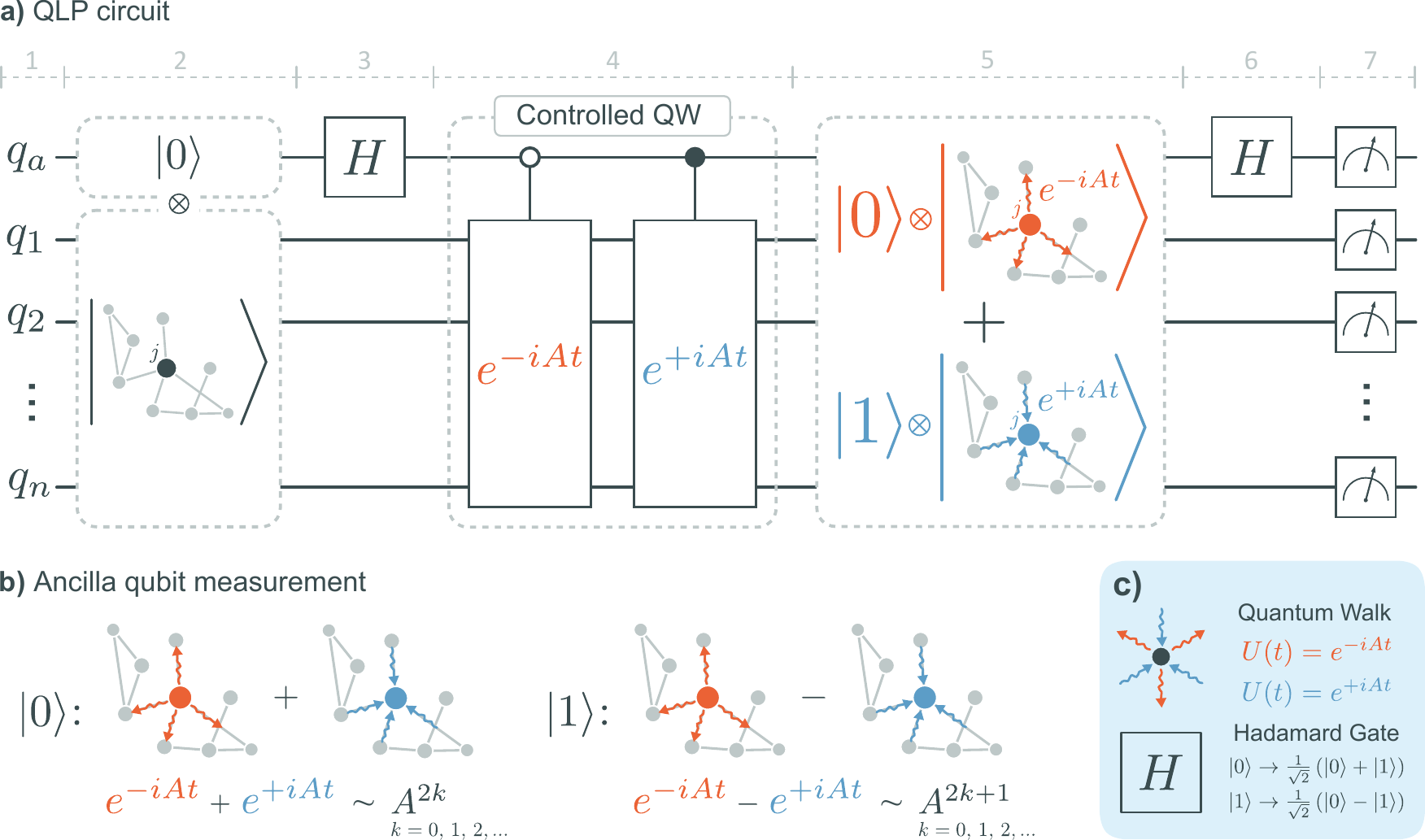}
\caption{\textbf{Quantum link prediction (QLP) circuit.} {\textbf{a)} 1: The algorithm requires a total of $n=\log_2(N)$ qubits to encode each of the $N$ nodes as a basis state, and an extra ancilla qubit $q_a$ to perform controlled operations. 2: The ancilla qubit is initialized to $\ket{0}$, and the remaining $n$ qubits are initialized to some basis state $\ket{j}$ corresponding to a node in the network. 3: The first Hadamard gate creates a superposition of $\ket{0}$ and $\ket{1}$ in $q_a$. 4: A controlled QW, represented here by two operators, applies $e^{-iAt}$ to the remaining $n$ qubits if $q_a=\ket{0}$ and $e^{+iAt}$ if $q_a=\ket{1}$.
5: Given that $q_a$ is in a superposition of $\ket{0}$ and $\ket{1}$, the controlled QW creates a superposition of the two evolutions entangled to the ancilla qubit.
6: A second Hadamard gate applied to the ancilla qubit mixes the two subspaces together and creates an interference between the two quantum walks.
7: Finally, all qubits are measured.
\textbf{b)} The measurement of $q_a$ collapses the network to one of two possible cases, imposing either a sum or subtraction of the two conjugate evolutions, which encodes even powers of $A$ (even predictions) for $q_a=\ket{0}$ and odd powers of $A$ (odd predictions) for $q_a=\ket{1}$ (Eq. \ref{eq:qlpstate0} and \ref{eq:qlpstate}). The measurement of the remaining $n$ qubits returns a bit string marking a node $i$, which together with the initial node $j$ forms a sample of a link $(i,\,j)$ (Eq. \ref{eq:qlpscores}). \textbf{c)} Legend for panels a) and b).}}
\label{fig:2}
\end{figure*}

\section{Quantum Link Prediction}
\label{sec:quantum}

We now describe our method for quantum link prediction, denoted as QLP, which we summarize at the end. We base our approach on a Continuous-Time Quantum Walk (CTQW) \cite{farhi1998quantum, kempe2003quantum}, where the Hilbert space of the quantum walker is defined by the orthonormal basis set $\{\ket{j}\}_{j\in\mathcal V}$, with each $\ket{j}$ corresponding to a localized state at a node $j$. We consider the Hamiltonian of the evolution as the adjacency matrix of the graph, $A$. In Figure \ref{fig:2} we show the main structure of the QLP circuit using a qubit representation. In the simplest case, we require $n=\log_2{N}$ qubits to add a binary label to each of the $N$ nodes, hereafter marked by the subscript $n$, and we consider an extra ancilla qubit $q_a$ that doubles the Hilbert space of the quantum walk, such that any node $j$ has two associated basis states,
\begin{equation}
\ket{0}_a\ket{j}_n~\text{and}~\ket{1}_a\ket{j}_n.
\end{equation}
For an initial state $\ket{\psi_j(0)}=\ket{0}_a\ket{j}_n$, the first step in the circuit of Figure \ref{fig:2} is to apply an Hadamard gate to $q_a$, which creates the superposition 
\begin{equation}
\frac{1}{\sqrt{2}}(\ket{0}+\ket{1})_a\ket{j}_n.
\end{equation}
A conditional CTQW is then applied which evolves the $q_a=\ket{0}$ subspace with $e^{-iAt}$ and the $q_a=\ket{1}$ subspace with $e^{+iAt}$. Finally, a second Hadamard gate is applied to $q_a$ to interfere the two quantum walks in the computational basis, leading to the state
\begin{equation}
\begin{aligned}
\ket{\psi_j(t)}=&\ket{0}_a\left(\frac{e^{-iAt}+e^{iAt}}{2}\right)\ket{j}_n\\
+&\ket{1}_a\left(\frac{e^{-iAt}-e^{iAt}}{2}\right)\ket{j}_n.
\end{aligned}
\label{eq:qlpstate0}
\end{equation}
To make the connection with link prediction more evident, we rewrite the previous expression as
\begin{equation}
\begin{aligned}
\ket{\psi_j(t)}=&\ket{0}_a\left(\sum_{k=0}^{+\infty}c_\text{even}(k, t)\,A^{2k}\right)\ket{j}_n\\
+i&\ket{1}_a\left(\sum_{k=0}^{+\infty}c_\text{odd}(k, t)\,A^{2k+1}\right)\ket{j}_n,
\end{aligned}
\label{eq:qlpstate}
\end{equation}
where we have replaced the exponential terms with their respective power series, and defined the time-dependent coefficients as
\begin{align}
c_\text{even}(k, t)&=(-1)^kt^{2k}/(2k)!\\
c_\text{odd}(k, t)&=(-1)^{k+1}t^{2k+1}/(2k+1)!.
\end{align}
A detailed calculation leading to Eq. \ref{eq:qlpstate} can be found in SI Section I. Given some initial node $j$, Eq. \ref{eq:qlpstate} describes the state that is created following the QLP circuit, before measurement. This state has two entangled components, one with a linear combination of even powers of $A$ for $q_a=\ket{0}$, and another with odd powers of $A$ for $q_a=\ket{1}$. The time $t$ of the quantum walk defines the linear weights, and acts as a hyperparameter in the model. This describes the unitary part of the protocol. To obtain relevant predictions from this state we must perform repeated measurements on the system to draw multiple samples, as we now describe.

The first step is to measure $q_a$, yielding $\ket{0}$ or $\ket{1}$ and collapsing the state of the remaining qubits to 
\begin{align}
\ket{\psi_{j}(t)}_n^\text{even}&\propto\left(\sum_{k}c_\text{even}(k, t)\,A^{2k}\right)\ket{j}~\text{or}\\
\ket{\psi_{j}(t)}_n^\text{odd}&\propto\left(\sum_{k}c_\text{odd}(k, t)\,A^{2k+1}\right)\ket{j},
\end{align}
respectively, where we omitted the normalization. This effectively selects whether the link sampled will be drawn from a distribution encoding even or odd powers of $A$. The last step is then to measure the remaining qubits, yielding a bit string corresponding to a sample of some node $i$ with probability
\begin{equation}
\begin{aligned}
&p_{ij}^\text{even}\propto\left|\mel{i}{\left(\sum_{k=0}^{+\infty}c_\text{even}(k, t)\,A^{2k}\right)}{j}\right|^2,\\
&p_{ij}^\text{odd}\propto\left|\mel{i}{\left(\sum_{k=0}^{+\infty}c_\text{odd}(k, t)\,A^{2k+1}\right)}{j}\right|^2,
\end{aligned}
\label{eq:qlpscores}
\end{equation}
which together with the initial node $j$ forms a sample of a link $(i,\,j)$. The values $p_{ij}^\text{even}$ and $p_{ij}^\text{odd}$ encode the prediction scores of the link $(i,\,j)$, but these can not be directly extracted from the algorithm. Instead, what this algorithm allows is the repeated sampling of these distributions, yielding pairs of nodes $(i,\,j)$ with probability proportional to $p_{ij}^\text{even}$ or $p_{ij}^\text{odd}$. This is analogous to sampling entries $(i,\,j)$ from the matrix of prediction scores $P$ with probability proportional to $|P_{ij}|^2$. As discussed in Section \ref{sec:clp}, predictions coming from even or odd powers of $A$ are typically useful in different types of networks. For a given network application of QLP, whether each sample obtained corresponds to an even or odd prediction depends on the value measured in the ancilla qubit, and this postselection can only be done probabilistically \cite{kothari2014efficient}. This is a potential sampling overhead, as unwanted predictions need to be discarded. Another overhead to consider is the possibility of sampling the initial node, or to sample already existing links, given the contribution of the identity $I$ in $p_{ij}^\text{even}$ and $A$ in $p_{ij}^\text{odd}$, which must also be discarded. As stated, QLP uses a linear combination of powers of $A$ weighted by the time $t$. As already mentioned, a classical prediction method with a linear combination of odd powers of $A$ was presented in \cite{pech2019link}, which was shown to sometimes improve the prediction precision compared to the original L3 method from \cite{kovacs2019network} by also fitting an additional model parameter. Another popular link prediction method is the Katz index \cite{katz1953new}, which uses a linear combination of all powers of $A$.

\begin{figure}
\centering
\includegraphics[width=\columnwidth]{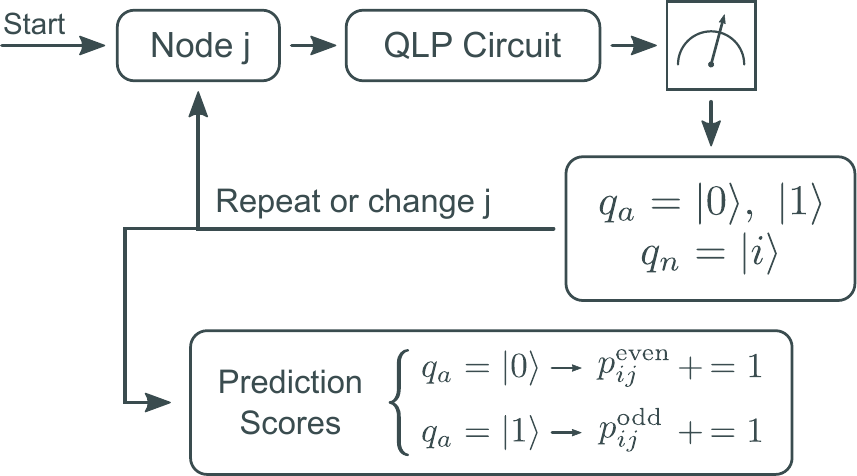}
\caption{\textbf{Quantum link prediction (QLP) algorithm.} Starting with an initial node $j$, the QLP circuit samples a node $i$ corresponding to an even or odd prediction of a link $(i,\,j)$ according to the value measured in $q_a$. Repeating this procedure for each node $j$ allows the larger values of the even and odd predictions scores $p_{ij}$ to be approximated.}
\label{fig:3}
\end{figure}

We can now summarize the QLP algorithm, as illustrated in Figure \ref{fig:3}. Firstly, an initial state $\ket{\psi_j(0)}=\ket{0}_a\ket{j}_n$ is prepared for a node $j$ in the network. Secondly, the QLP evolution leading to Eq. \ref{eq:qlpstate} is performed for a specific time $t$. Finally, the ancilla and node qubits are measured to obtain a sample of a link $(i,j)$ corresponding to an even or odd prediction, and the procedure is repeated. The number of samples that output a certain link $(i,j)$ will follow the distributions described by Eq. \ref{eq:qlpscores}, and thus represent a score for link $(i,j)$. Once predictions associated with node $j$ are sufficiently characterized, the procedure can be repeated for other relevant nodes in the network.

\section{Results and Discussion}
\label{sec:results}
\subsection{Complexity analysis}

To identify a potential quantum advantage, we briefly discuss how link prediction scales on a classical computer. Complex networks are typically sparse \cite{barabasi2016network} with the average degree much smaller than the total number of nodes, $k_\mathrm{av}\ll N$, and as such there are $\mathcal{O}(N^2)$ potentially missing links. The general case of computing all possible scores leads to a classical complexity of at least $\mathcal{O}(N^2)$. Different methods scale differently depending on the assumptions made about the solution. For example, the scaling of simple length-2 based methods is $\mathcal{O}(N\langle k^2\rangle)$ and the scaling of L3 \cite{kovacs2019network} is upper bounded by $\mathcal{O}(N\langle k^3\rangle)$, where $\langle k^n\rangle$ is the average of the $n$-th power of the degrees (see SI Section II). These methods do not calculate a score for every possible missing link, only for those corresponding to nodes at distance 2 or 3. However, other methods also surpass the $\mathcal{O}(N^2)$ scaling, as is the case of LO \cite{pech2019link} that uses a matrix inversion to represent a linear combination of odd powers of $A$, scaling approximately with $\mathcal{O}(N^{2.4})$, and is one of the best performing classical methods tested. Complex networks can easily reach sizes of up to millions or billions of nodes, consider for example online social and e-commerce networks, or the neuronal network in the human brain \cite{azevedo2009equal}. Improving these scalings may thus be decisive in the application of link prediction methods to larger networks in the future.

To provide an estimate for the complexity of implementing QLP on a quantum computer, there are a few things to consider. First, we comment on the implementation of the $e^{-iAt}$ unitary, representing the CTQW used to obtain each link prediction sample. For this purpose, the most relevant results are related to the quantum simulation of $d$-sparse matrices, meaning that $A$ has at most $d$ entries in any given row. A state of the art result \cite{low2017optimal} shows that in that scenario implementing $e^{-iAt}$ scales as $\tilde{\mathcal{O}}\left(dt\|A\|_\text{max}\right)$, omitting logarithmic factors, where $t$ is the time interval of the evolution and $\|A\|_\text{max}$ is the maximum entry in absolute value. In our case, $d=k_\text{max}$, the maximum degree of the network, and $\|A\|_\text{max}=1$, which allows us to write the complexity of implementing $e^{-iAt}$ as $\mathcal{\tilde{O}}(k_\text{max}t)$. Second, we comment on the time $t$. As mentioned earlier, $t$ is a hyperparameter in the model which determines how each power of $A$ is weighted for the predictions. A large value of $t$ would lead higher powers of $A$ to be more heavily weighted. This is not the case for typical link prediction methods, where the most relevant contributions are typically from $A^2$ or $A^3$, irrespectively of the network size. In our simulations we found the optimal value of $t$ to change from network to network, however it seems to do so independently of $N$ (see SI Table 2). For these reasons, we believe it is reasonable to disregard the contribution of $t$ to the complexity. Finally, for each application of the circuit from Fig. \ref{fig:2} a link prediction sample associated with a node $j$ is obtained. Then, assuming a repetition of the process to obtain samples for every node leads to a factor of $N$ in the complexity, and for each node $j$ a sufficient number of samples $s_j$ is required to characterize the predictions associated with it. If we consider that the missing links have been removed randomly from the network, each node $j$ will have a number of missing links proportional to its observed degree $k_j$. However, the relation between $s_j$ and $k_j$ is highly network-dependent, as it depends on how well the quantum walk method represents the underlying truth of the missing links. In practice, we can leave $s_j$ as a free parameter, as ultimately the number of samples to obtain would be decided by the user. 

In summary, assuming a repetition of the QLP method for each node in a network of $N$ nodes, with an average number of samples per node of $s_\text{av}=\frac{1}{N}\sum_{j=1}^N s_j$, and each sample requiring the implementation of $e^{-iAt}$ with a cost $\tilde{O}(k_\text{max})$ for some constant $t$, the final complexity estimate for QLP is
\begin{equation}
    \mathcal{\tilde{O}}(Ns_\text{av}k_\text{max}).
\end{equation}

The most meaningful complexity comparison we can make is between methods that make similar assumptions. In that sense, both QLP and LO assume the solution is a linear combination of powers of the adjacency matrix, and as we will see in the next section, these methods are often the best performing. Here, we can see that QLP has a potential quantum speedup given the polynomially lower dependence on $N$ but with an extra $s_\text{av}k_\text{max}$ factor. Relating $k_\text{max}$ to $N$ can be done through $\gamma$ as $k_\text{max}\propto N^{\frac{1}{\gamma-1}}$, where $\gamma$ is the exponent in the power-law degree distribution of a scale-free network, which is typically in the range $2<\gamma\leq 4$ \cite{barabasi2016network}. For these values the dependence is always sub-linear, approaching linearity as $\gamma\rightarrow 2$, as in this regime the network tends to form larger and larger hubs. For $\gamma=3$, for example, our estimate for the scaling of QLP in scale-free networks is $\mathcal{\tilde{O}}(N^{3/2}s_\text{av})$, a potential polynomial speedup over the LO method. Comparing QLP to simple length-2 and length-3 based methods is less straightforward, as the difference is solely based on the degree factors and the number of samples for QLP.

\subsection{Cross-validation tests}

\begin{figure*}[t]
\centering
\includegraphics[width=1.0\textwidth]{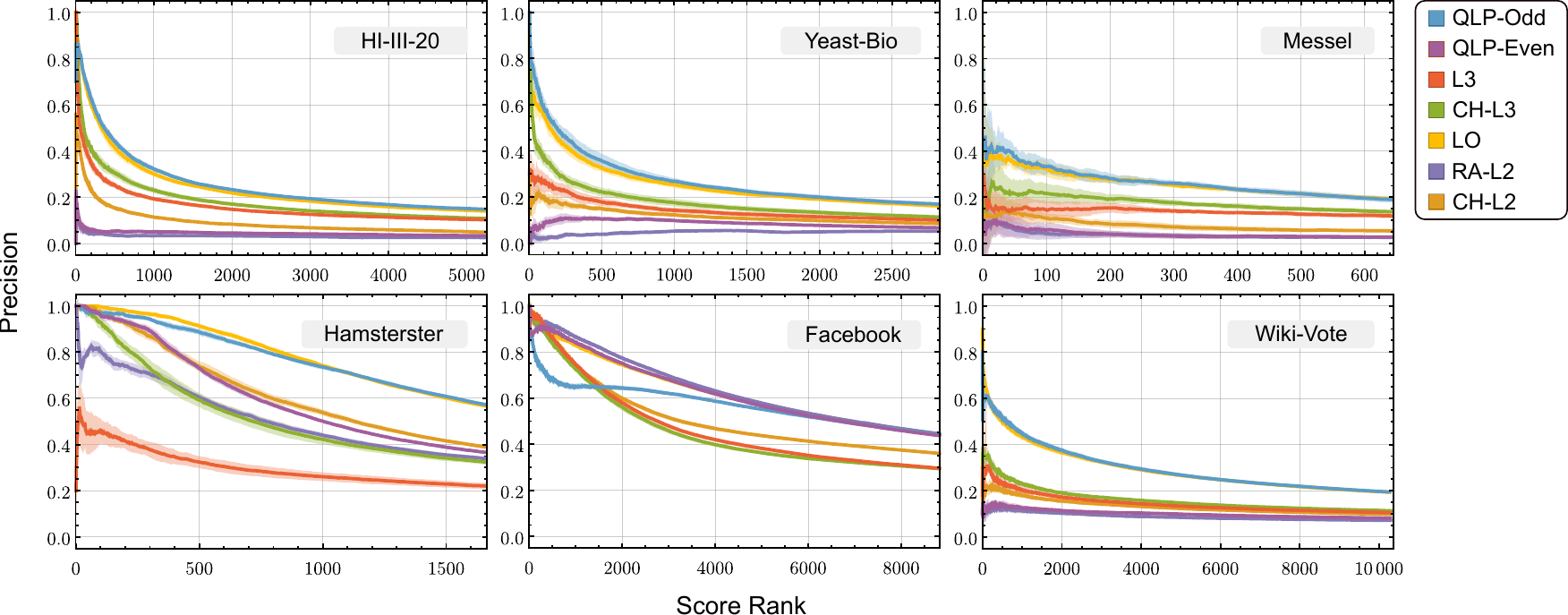}
\caption{\textbf{Computational cross-validation: Precision.} Cumulative precision over the list of ranked scores for each network, averaged over a 10-fold cross validation procedure. The shaded regions correspond to the standard deviation. Details on the Precision and Score Rank metrics are shown in SI Section III. The networks used correspond to the PPI networks HI-III-20, the most recent PPI mapping of the human interactome \cite{luck2020interactome}, Yeast-Bio, a PPI network of a yeast organism \cite{biogrid}, Messel, a food web \cite{messel}, Hamsterster \cite{hamsterster} and Facebook \cite{facebook}, two online social networks, and Wiki-Vote, a vote network between users for adminship of Wikipedia \cite{wikivote}. For comparison, we implemented five classical link prediction methods: the L3 method \cite{kovacs2019network}, the LO method \cite{pech2019link}, the CH-L3 method \cite{Cannistraci:2018}, and two even power methods, RA-L2 (resource allocation) \cite{Zhou:2009}, and CH-L2 \cite{cannistraci2013link, Cannistraci:2018}. The dataset parameters characterizing each network are shown in SI Table I, and the values selected for the optimal parameters $t$ in the QLP method and $\alpha$ in the LO method are shown in SI Table II.}
\label{fig:4}
\end{figure*}

\begin{table*}[t]
\footnotesize
\centering
\begin{tabular}{cc|cc|ccccc}
& Dataset  & QLP-Even & QLP-Odd & LO & L3 & CH-L3 & RA-L2 & CH-L2  \\\hline
\multirow{6}{0.1cm}{\rotatebox{90}{\textbf{AUC-ROC}}}
&HI-III-20 & 0.786 & \textbf{0.909} & 0.879 & \textbf{0.917} & 0.917 & 0.655 & 0.655 \\
&Yeast-Bio & 0.878 & \textbf{0.894} & 0.852 & \textbf{0.905} & 0.904 & 0.738 & 0.738 \\
&Messel    & 0.635 & \textbf{0.887} & 0.880 & \textbf{0.891} & 0.890 & 0.641 & 0.649 \\
&Hamsterster & \textbf{0.971} & 0.964 & 0.952 & 0.965 & \textbf{0.966} & 0.962 & 0.962 \\
&Facebook  & \textbf{0.995} & 0.994 & 0.988 & 0.991 & 0.991 & \textbf{0.995} & 0.994 \\
&Wiki-Vote & 0.878 & \textbf{0.904} & 0.898 & \textbf{0.905} & \textbf{0.905} & 0.858 & 0.859 \\\hline
\multirow{6}{0.1cm}{\rotatebox{90}{\textbf{AUC-PR}}}
&HI-III-20 & 0.006 & \textbf{0.081} & \textbf{0.074} & 0.042 & 0.049 & 0.005 & 0.013 \\
&Yeast-Bio & 0.014 & \textbf{0.093} & \textbf{0.082} & 0.038 & 0.049 & 0.013 & 0.024 \\
&Messel    & 0.008 & \textbf{0.104} & \textbf{0.104} & 0.051 & 0.062 & 0.007 & 0.013 \\
&Hamsterster & 0.341 & \textbf{0.568} & \textbf{0.574} & 0.131 & 0.280 & 0.284 & 0.365 \\
&Facebook  & \textbf{0.429} & 0.392 & 0.427 & \textbf{0.444} & 0.334 & 0.262 & 0.257 \\
&Wiki-Vote & 0.0287 & \textbf{0.112} & \textbf{0.111} & 0.026 & 0.037 & 0.043 & 0.047 \\\hline
\end{tabular}
\caption{\textbf{Computational cross-validation: AUC.} Area under the curve (AUC) performance metrics for the datasets in Figure \ref{fig:4}. The metrics were computed over the full set of potential predictions for each dataset in a 10-fold cross-validation procedure. Each value corresponds to the mean AUC over the ten iterations. In bold are the best values for both QLP and the classical methods tested, confirming that QLP performs similarly to other path-based link prediction methods in the AUC-ROC and AUC-PR metrics.}
\label{tab:auc}
\end{table*}

In Figure \ref{fig:4} we compare the prediction precision of QLP with classical path-based link prediction methods using the standard link prediction benchmark of cross-validation on a selection of networks from different fields. In Table \ref{tab:auc} we further compare the methods using two standard AUC metrics. A summarized description of the classical methods used can be found in Ref.\ \cite{zhou2021progresses}. We compared against three odd-power methods, L3 \cite{kovacs2019network}, LO \cite{pech2019link} and CH-L3 \cite{Cannistraci:2018}, and two even-power methods, RA-L2 \cite{Zhou:2009} and CH-L2 \cite{cannistraci2013link, Cannistraci:2018}. These are the state-of-the-art in local and global link prediction indices based on path-counting. The scores used for QLP were an exact calculation of the distributions in Eq. \ref{eq:qlpscores} by classically computing the time-evolution operator of the quantum walk, as described in SI Section I.A. For each network, we selected the time $t$ that maximizes the prediction precision by removing 10\% of the links from the training set in the first iteration of the cross-validation, selecting both a value that maximizes the precision of the even component as well as one that maximizes the odd component, detailed in SI Table II. As shown in Figure \ref{fig:4} and Table \ref{tab:auc}, we confirm that QLP matches the typical performance of classical path-based link prediction methods tested in terms of prediction precision as well as standard AUC metrics for a range of real life complex networks \cite{luck2020interactome, biogrid, messel, hamsterster, facebook, wikivote}, as expected. In most cases, we observe that both QLP-Odd and LO stand out as the best performing methods, a result which further affirms the case that there can be advantages in including higher order powers of the adjacency matrix in the predictions \cite{pech2019link}. Further results for the cross-validation benchmark are shown in SI Figure 1, as well as detailed results for each of the experimental screens that contribute to the full HI-III-20 network in SI Figure 2. Here, we predict interactions that have been obtained by independent, full experimental screens, simulating the case of real life performance against future experiments.

\section{Conclusions}

In this work we have presented a quantum algorithm for link prediction in complex networks, QLP, offering a potential quantum speedup for a practical network science problem. The inclusion of even and odd paths allows QLP to make both TCP-like and L3-like predictions, thus encompassing all types of networks where these topological patterns play a role. Our results serve as a proof of principle for potential future applications of QLP in large complex networks using quantum hardware. Recently, a 62-node network CTQW was demonstrated experimentally \cite{gong2021quantum}, an important first step towards this goal. 

We note that, in our estimated complexity, the dependence on $k_\text{max}$ comes from assuming a circuit-based simulation of the quantum walk with the $d$-sparse matrix model. We may argue that the existence of large hubs make complex networks a bad fit for the $d$-sparse matrix model. Highly connected nodes imply that some rows in the adjacency matrix are very dense, while most are sparse, and thus $d=k_\text{max}$ greatly overestimates the overall sparseness of the matrix. Finding a more efficient quantum simulation algorithm that directly exploits the degree distribution of complex networks would be a very important result for quantum computation applied to network science, should such a method exist. Nevertheless, the method we propose here is general to any representation of the quantum walk, for example using an analog quantum walk implementation, as done in \cite{gong2021quantum}, or any future quantum simulation techniques that prove to be more efficient.

Besides the potential improvement in complexity when sampling from the quantum solution, especially in the comparison between QLP and LO, we should also note that a classical simulation of QLP relies on the diagonalization of the adjacency matrix, and thus it has a comparable classical complexity to other path-based classical link prediction methods. This makes QLP easier to be further developed with a focus on immediately relevant real world applications, while at the same time exploring other ways in which quantum features of QLP can be advantageous when quantum hardware becomes more widely available.

\section*{Code Availability}
\small
Our code for QLP is available at\\
\href{https://github.com/jpmoutinho/Quantum-Link-Prediction}{https://github.com/jpmoutinho/Quantum-Link-Prediction}.

\section*{Acknowledgements}
\small
The authors thank Albert-László Barabási for the useful discussion, and acknowledge the support from the JTF project \textit{The Nature of Quantum Networks} (ID 60478). JPM, BC and YO thank the support from Funda\c{c}\~{a}o para a Ci\^{e}ncia e a Tecnologia (FCT, Portugal), namely through projects UIDB/50008/2020 and UIDB/04540/2020, as well as from projects TheBlinQC and QuantHEP supported by the EU H2020 QuantERA ERA-NET Cofund in Quantum Technologies and by FCT (QuantERA/0001/2017 and QuantERA/0001/2019, respectively), and from the EU H2020 Quantum Flagship project QMiCS (820505). JPM acknowledges the support of FCT through scholarship SFRH/BD/144151/2019, and BC acknowledges the support of FCT through project CEECINST/00117/2018/CP1495.

\bibliographystyle{apsrev4-1}
\bibliography{manuscript}

\clearpage

\setcounter{figure}{0}
\setcounter{table}{0}
\onecolumngrid

\begin{center}
\large
\textbf{Supplementary Information for}\\
\textbf{Quantum Link Prediction in Complex Networks}
\end{center}

\section{QLP method}

We consider the usual continuous-time quantum walk (CTQW) model, where the Hilbert space of the quantum walker is defined by the orthonormal basis set $\{\ket{j}\}_{j\in\mathcal V}$, each basis state $\ket{j}$ corresponding to a localized state at a node $j$ in the network, and the Hamiltonian of the evolution given by the adjacency matrix of the graph. In these conditions, the solution to the Schr\"{o}dinger equation for the CTQW can be written directly as
\begin{equation}
\ket{\psi(t)}=e^{-iAt}\ket{\psi(0)}.
\end{equation}
By taking the power series of the time evolution operator we can immediately make the connection to link prediction,
\begin{equation}
e^{-iAt}=\sum_{k=0}^{+\infty}\frac{1}{k!}(-it)^{k}A^{k}, \label{eq:powerseries}
\end{equation}
as each power $A^k$ encodes the number of paths of length $k$ between any two nodes in the graph. Furthermore, we note that the imaginary term $i^k$ adds a phase to the quantum evolution that separates the sum over even powers in the real part of the evolution and the sum over odd powers in the imaginary part. To proceed we wish to separate the evolution over even powers from the evolution over odd powers, and for that it is useful to consider a qubit representation of the graph, as seen in Fig. 2 of the main text. We now define an operator $\text{CQW}(t)$ corresponding to a controlled quantum walk which applies a normal or conjugate evolution operator on the node qubits depending on the value of $q_a$,
\begin{equation}
\text{CQW}(t)=\ketbra{0}{0}_a\left(e^{-iAt}\right)_n+\ketbra{1}{1}_a\left(e^{+iAt}\right)_n.
\end{equation}
Considering now an initial state localized at node $j$ in the form $\ket{\psi_j(0)}=\ket{0}_a\ket{j}_n$ we start by applying an Hadamard gate on the ancilla qubit,
\begin{equation}
H_a\ket{0}_a\ket{j}_n = \frac{1}{\sqrt{2}}(\ket{0}+\ket{1})_a\ket{j}_n,
\end{equation}
followed by the $\text{CQW}(t)$ operator,
\begin{equation}
\text{CQW}(t)\left[\frac{1}{\sqrt{2}}(\ket{0}+\ket{1})_a\ket{j}_n\right]=\frac{1}{\sqrt{2}}\left(\ket{0}_ae^{-iAt}\ket{j}_n+\ket{1}_ae^{+iAt}\ket{j}_n\right),
\end{equation}
followed by a second Hadamard gate on the ancilla qubit, leading to the following expression after rearranging the terms:
\begin{equation}
\ket{\psi_j(t)}=\frac{1}{2}\ket{0}_a\left(e^{-iAt}+e^{iAt}\right)\ket{j}_n + \ket{1}_a\left(e^{-iAt}-e^{iAt}\right)\ket{j}_n. \label{eq:qlpstate1}
\end{equation}
Finally, taking the power series from Eq. \ref{eq:powerseries} to replace the exponential terms we arrive at 
\begin{equation}
\ket{\psi_j(t)}=\ket{0}_a\left(\sum_{k=0}^{+\infty}c_\text{even}(k, t)\,A^{2k}\right)\ket{j}_n + i\ket{1}_a\left(\sum_{k=0}^{+\infty}c_\text{odd}(k, t)\,A^{2k+1}\right)\ket{j}_n, \label{eq:qlpstate2}
\end{equation}
with $c_\text{even}(k, t)=(-1)^kt^{2k}/(2k)!$ and $c_\text{odd}(k, t)=(-1)^{k+1}t^{2k+1}/(2k+1)!$ being time-dependent coefficients.

This procedure describes how the real and imaginary part of the time-evolution operator can be separated on a quantum computer through an extra ancilla qubit. To simulate QLP on a conventional computer, it suffices to compute the time-evolution operator and directly extract the real and imaginary part, as described below.

\subsection{QLP on a conventional computer}

In this section we describe how to directly compute the scores of QLP on a conventional computer. Consider a network described by its adjacency matrix $A$. Start by picking a value for the $t$ parameter. Then, numerically compute the time-evolution operator of the quantum walk,
\begin{equation}
    U(t)=e^{-iAt},
\end{equation}
for example, by computing the eigenvalues and eigenvectors of $A$ and then computing the matrix exponential. The matrix $U(t)$ is complex. The prediction scores described in the main text can then be obtained directly, in matrix form, as
\begin{equation}
    \begin{aligned}
    P^\text{even}&=|\text{Re}(e^{-iAt})|_{(ij)}^2\\
    P^\text{odd}&=|\text{Im}(e^{-iAt})|_{(ij)}^2
    \end{aligned}
\end{equation}
where $|.|_{ij}^2$ denotes the entry-wise absolute value squared. The entries of $P^\text{even}$ and $P^\text{odd}$ correspond to the $p_{ij}^\text{even}$ and $p_{ij}^\text{odd}$ values described in the main text, which can be used to rank predictions from highest to lowest score.

\section{Link prediction complexity}

Consider a graph $G(\mathcal{V},\,\mathcal{E})$ describing a complex network, where $\mathcal{V}$ is the set of nodes with size $N=|\mathcal{V}|$ and $\mathcal{E}$ is the set of undirected links. Link prediction on a classical computer requires $\frac{1}{2}N(N-1)-|\mathcal{E}|$ scores to be computed, one for each of the $\frac{1}{2}N(N-1)$ possible links, with the exception of those already present in the set of known links $\mathcal{E}$. Rewriting in terms of the average degree, $k_\mathrm{av}=2|\mathcal{E}|/N$, we have that the total number of scores scales as $\frac{1}{2}N^2-\frac{1}{2}N(1+k_\mathrm{av})$. Real complex networks are typically sparse \cite{barabasi2016network} with $k_\mathrm{av}\ll N$, and thus $\mathcal{O}(N^2)$ scores are evaluated. Taking $\mathcal{O}(N^2)$ as an estimate for the complexity of a general classical link prediction method assumes two more things: that the method will indeed compute a score for every potential missing link, and that the cost of computing each score is $\mathcal{O}(1)$. In order to analyse these assumptions, let us pick a concrete method and study its complexity.

Common Neighbours (CN) is one of the simplest link prediction algorithms. It quantifies the likelihood of a link existing between two nodes $i$ and $j$ by the number of common neighbours they share, or in other words, by the number of paths of length 2 between $i$ and $j$. While we do not use CN directly in the various simulations presented in this work, we used the method of Resource Allocation \cite{Zhou:2009} (marked as RA-L2 in the plots), which is similar to CN with the addition of a degree normalization to each score. Adding the degree normalization does not affect the complexity significantly, and so we will analyse the simpler problem of counting paths of length 2. The objective of CN is to compute
\begin{equation}
p_{ij}=|\Gamma(i)\cap\Gamma(j)|
\end{equation}
for every pair of nodes $(i,j)$ where $|\Gamma(i)\cap\Gamma(j)|\neq 0$, $\Gamma(x)$ being the set of nodes neighbouring $x$. A simple algorithm to accomplish this iterates through all nodes $z$ in the graph and adds a contribution to $p_{ij}$ for each pair of nodes $(i,j)$ neighbouring $z$. Such an algorithm will visit every path of length 2 in the graph and thus its complexity will be proportional to $\sum_{i,j=1}^N(A^2)_{ij}$. As detailed in \cite{fiol2009number} this sum can be simplified as
\begin{equation}
\sum_{i,j=1}^N(A^2)_{ij}=\sum_{i=1}^Nk_i^2=N\langle k^2\rangle, 
\end{equation}
where $\langle k^2\rangle$ is the average of the second power of the degrees in the graph. By assuming that the cost of accessing the graph data structure and adding the contributions to each $p_{ij}$ is $\mathcal{O}(1)$ we can conclude that the CN method scales as $\mathcal{O}(N\langle k^2\rangle)$.

Common Neighbours is a TCP based method, and as discussed in the main text, it is not able to match the precision of methods based on paths of length 3 in many networks. For that reason, let us see how the complexity changes when counting paths of length 3, which is the main computational cost behind the L3 method \cite{kovacs2019network}. An algorithm to count paths of length 3 can be easily built with an extension of the CN algorithm, and using the same argument as before, its complexity will be proportional to $\sum_{i,j=1}^N(A^3)_{ij}$. This sum is not as easy to simplify, but the authors in \cite{fiol2009number} prove the following bound for a general power of $A$
\begin{equation}
\sum_{i,j=1}^N(A^n)_{ij}\leq\sum_{i=1}^Nk_i^n=N\langle k^n\rangle.
\end{equation}
With this information we can conclude that the complexity for counting paths of length 3 will be upper bounded by $\mathcal{O}(N\langle k^3\rangle)$.

For scale-free networks, characterized by $\gamma$, the degree exponent in the degree power law distribution, we can analyse the moments $\langle k^n\rangle$ in terms of $\gamma$ and $N$ (see Section 4 of \cite{barabasi2016network}). Typically, $\langle k\rangle$ (denoted as $k_\text{av}$ in the rest of the text) is much smaller than $\langle k^2\rangle$ or $\langle k^3\rangle$. For many scale-free networks $\gamma$ is between 2 and 4. As $N$ grows, $\langle k^2\rangle$ diverges for $2<\gamma\leq 3$ and $\langle k^3\rangle$ diverges for $2<\gamma\leq 4$, while $\langle k\rangle$ does not. These divergences can be seen in the expressions below from \cite{barabasi2016network} which estimate the dependence of $\langle k^n \rangle$ with $N$
\begin{equation}
\langle k^n\rangle\propto\frac{k_\text{max}^{n-\gamma+1}-k_\text{min}^{n-\gamma+1}}{n-\gamma+1}
\end{equation}
which together with the relation $k_\text{max}=k_\text{min}N^\frac{1}{\gamma-1}$ can be written as
\begin{equation}
\langle k^n\rangle\propto \frac{k_\text{min}^{n-\gamma+1}}{n-\gamma+1}\left(N^\frac{n-\gamma+1}{\gamma-1}-1\right)
\end{equation}

Out of the methods tested in this work, RA-L2 and L3 fall in the complexity categories of counting paths of length 2 and 3, respectively. CH-L2 and CH-L3 also have path counting as a base, but use a more complex structure of paths which has added complexity. LO, the best performing classical method tested, uses a matrix inversion for which the best algorithms scale roughly as $\mathcal{O}(N^{2.4})$.

As stated in the main text, the complexity of QLP can be written as $\mathcal{\tilde{O}}(Nk_\text{av}k_\text{max}t)$. The previous expressions show that $k_\text{av}$ remains finite for all $\gamma > 2$, while higher order moments can diverge. Although these expressions do not include any information about the finite value to which the moments tend when they do not diverge, complex networks are typically sparse, we may still use $k_\text{av}\ll N$ to quantify the differences in complexity between the methods, especially in the cases where $\langle k^2\rangle$ and $\langle k^3\rangle$ diverge with growing $N$. Furthermore, we can comment on the dependence with $k_\text{max}$ coming from the $d$-sparse Hamiltonian simulation of the quantum walk. The relation $k_\text{max}\propto N^\frac{1}{\gamma-1}$ \cite{barabasi2016network} leads to $k_\text{max}\propto N$ in the limit of $\gamma\rightarrow 2$, implying a quadratic scaling of QLP. This lower bound corresponds to an extreme case in scale-free networks, and other larger values within the realistic $2<\gamma<4$ range reduce this dependence on $N$ polynomially.

\section{Precision and AUC Metrics}
In the main text precision plot we show the cumulative precision over the score rank for each method and network tested. The score rank represents the ordered list of scores: the top score has score rank 0, the second best has score rank 1, and so on. The cumulative precision tracks the ratio of correct predictions to total predictions over all previous score ranks. For example, a precision of 0.8 at score rank 9 means that out of the 10 top predictions occupying score rank 0 through 9, eight of them were correct. For each iteration of the 10-fold cross validation procedure 10\% of the links were randomly removed and the remainder used as input to the link prediction methods, leading to a different cumulative precision curve for each iteration. In the plots we show the average cumulative precision $\pm$ one standard deviation over the 10 iterations. Since the networks tested have different sizes and densities the total number of predictions that may be considered relevant varies. We chose to cutoff the figure at a score rank of $0.05\times N\times k_\text{av}$ which is sufficient to show a drop in the precision over the score rank while still focusing on the precision of the best predictions occupying the first ranks.

In the main text AUC table we show the AUC-ROC and AUC-PR. These are the areas under the receiver-operating characteristic curve and precision-recall curve, respectively, which are standard benchmark metrics for predictive models.

\clearpage
\begin{table}[t]
\centering
\caption{Datasets and respective network parameters. \label{table:datasetparameters}}
\begin{tabular}{cccccccccc}
& Network  & Ref. & $|V|$ & $|E|$ & $k_\text{av}$ & $\rho$ & $d_\text{max}$ & $d_\text{av}$ & $C$ \\\hline
\multirow{6}{0.5cm}{\rotatebox{90}{Main Text - Fig. 4}}
& HI-III-20 &\cite{luck2020interactome}& 8275 & 52569 & 12.589 & $1.59\times 10^{-3}$ & 12 & 3.844 & $5.92\times 10^{-2}$ \\
& Yeast-Bio &\cite{biogrid}& 4885 & 28270 & 11.161 & $2.29\times 10^{-3}$ & 10 & 3.603 & $1.20\times 10^{-1}$ \\
& Messel &\cite{messel}& 700 & 6444 & 18.326 & $2.61\times 10^{-2}$ & 6 & 2.632 & $1.04\times 10^{-1}$\\
& Hamsterster &\cite{hamsterster}& 2426 & 16631 & 13.711 & $5.65\times 10^{-3}$ & 10 & 3.589 & $5.38\times 10^{-1}$ \\
& Facebook &\cite{facebook}& 4039 & 88234  & 43.691 & $1.08\times 10^{-2}$ & 8 & 3.693 & $6.06\times 10^{-1}$ \\
& Wiki-Vote &\cite{wikivote}& 7115 & 103689 & 29.147 & $3.98\times 10^{-3}$ & 7 & 3.248 & $1.41\times 10^{-1}$ \\\hline
\multirow{6}{0.5cm}{\rotatebox{90}{S.I. - Fig. 1}}
& Arabidopsis &\cite{arabidopsis}& 4865 & 11374 & 4.493 & $9.24\times 10^{-4}$ & 14 & 5.180 & $9.82\times 10^{-2}$ \\
& Pombe &\cite{biogrid}& 1929 & 3700 & 3.397 & $1.76\times 10^{-3}$ & 14 & 4.671 & $6.37\times 10^{-2}$ \\
& AS Routes &\cite{asroutesp2p}& 6474  & 13895 & 3.884 & $6.00\times 10^{-4}$ & 9 & 3.705 & $2.52\times 10^{-1}$ \\
& Citeseer &\cite{citeseercora}& 3264 & 4536 & 2.779 & $8.518\times 10^{-4}$ & 28 & 9.315 & $1.45\times 10^{-1}$ \\
& Cora &\cite{citeseercora}& 2708 & 5429 & 4.010 & $1.44\times 10^{-3}$ & 19 & 6.311 & $2.41\times 10^{-1}$ \\
& P2P-Gnutella &\cite{asroutesp2p}& 10876 & 39994 & 7.355 & $6.78\times 10^{-4}$ & 10 & 4.622 & $6.22\times 10^{-3}$ \\\hline
\multirow{9}{0.5cm}{\rotatebox{90}{S.I. - Fig. 2 - HuRI Screens}}
& Screen 1 &\cite{luck2020interactome}& 4643 & 16447 & 6.970 & $1.64\times 10^{-3}$ & 12 & 4.094 & $5.13\times 10^{-2}$ \\
& Screen 2 &\cite{luck2020interactome}& 4177 & 11644 & 5.467 & $1.31\times 10^{-3}$ & 13 & 4.284 & $4.33\times 10^{-2}$ \\
& Screen 3 &\cite{luck2020interactome}& 3807 & 10245 & 5.268 & $1.31\times 10^{-3}$ & 15 & 4.456 & $4.16\times 10^{-2}$ \\
& Screen 4 &\cite{luck2020interactome}& 3082 & 5685  & 3.655 & $1.19\times 10^{-3}$ & 14 & 5.370 & $1.51\times 10^{-2}$ \\
& Screen 5 &\cite{luck2020interactome}& 2712 & 4496  & 3.277 & $1.21\times 10^{-3}$ & 15 & 5.560 & $1.25\times 10^{-2}$ \\
& Screen 6 &\cite{luck2020interactome}& 3128 & 5981  & 3.774 & $1.21\times 10^{-3}$ & 16 & 5.361 & $1.54\times 10^{-2}$ \\
& Screen 7 &\cite{luck2020interactome}& 3508 & 7910  & 4.486 & $1.28\times 10^{-3}$ & 15 & 5.465 & $9.73\times 10^{-3}$ \\
& Screen 8 &\cite{luck2020interactome}& 3383 & 7533  & 4.436 & $1.31\times 10^{-3}$ & 16 & 5.555 & $1.20\times 10^{-2}$ \\
& Screen 9 &\cite{luck2020interactome}& 3404 & 7712  & 4.512 & $1.33\times 10^{-3}$ & 15 & 5.520 & $9.40\times 10^{-3}$ \\
\hline
\end{tabular}
\end{table}

Network parameters: $|V|$ is the number of nodes; $|E|$ is the number of links; $k_\text{av}$ is the average degree; $\rho$ is the network density; $d_\text{max}$ is the maximum distance between any two nodes; $d_\text{av}$ is the average distance between any two nodes; $C$ is the average clustering coefficient.

\clearpage
\begin{table}[t]
\centering
\begin{tabular}{ccccccccc}
& Network  & $t$ (QLP-Even) & $t$ (QLP-Odd) & $\alpha$ (LO) \\\hline
\multirow{6}{0.5cm}{\rotatebox{90}{Main Text - Fig. 4}}
& HI-III-20 & $3.0\times 10^{-6}$ & 1.00 & $8.0\times 10^{-3}$ \\
& Yeast-Bio & $7.0\times 10^{-1}$ & 1.10 & $2.0\times 10^{-2}$ \\
& Messel    & $2.0\times 10^{-6}$ & 1.20 & $2.0\times 10^{-2}$ \\
& Hamsterster & $4.0\times 10^{-1}$ & 1.60 & $3.0\times 10^{-2}$ \\
& Facebook   & $2.0\times 10^{-1}$ & 1.00 & $7.0\times 10^{-3}$ \\
& Wiki-Vote  & $2.0\times 10^{-6}$ & 1.00 & $4.0\times 10^{-3}$ \\\hline
\multirow{6}{0.5cm}{\rotatebox{90}{S.I. - Fig. 1}}
& Arabidopsis & $2.0\times 10^{-6}$ & 1.00 & $2.0\times 10^{-2}$ \\
& Pombe       & $9.0\times 10^{-1}$ & 1.00 & $4.0\times 10^{-2}$ \\
& AS Routes  & $5.0\times 10^{-4}$ & 0.60 & $2.0\times 10^{-3}$ \\
& Citeseer   & $5.0\times 10^{-5}$ & 1.40 & $8.0\times 10^{-2}$ \\
& Cora       & $1.0\times 10^{-6}$ & 0.03 & $6.0\times 10^{-2}$ \\
& P2P-Gnutella & - & 1.30  & $2.0\times 10^{-2}$  \\\hline
\multirow{9}{0.5cm}{\rotatebox{90}{S.I. - Fig. 2 - HuRI Screens}}
& Screen 1 & $6.0\times 10^{-6}$ & 1.10 & $1.0\times 10^{-2}$ \\
& Screen 2 & $4.0\times 10^{-6}$ & 0.90 & $8.0\times 10^{-3}$ \\
& Screen 3 & $6.0\times 10^{-5}$ & 0.90 & $1.0\times 10^{-2}$ \\
& Screen 4 & $9.0\times 10^{-5}$ & 0.02 & $1.0\times 10^{-2}$ \\
& Screen 5 & $6.0\times 10^{-5}$ & 0.03 & $3.0\times 10^{-3}$ \\
& Screen 6 & $1.0\times 10^{-6}$ & 0.03 & $7.0\times 10^{-3}$ \\
& Screen 7 & $1.0\times 10^{-6}$ & 0.70 & $7.0\times 10^{-4}$ \\
& Screen 8 & $2.0\times 10^{-6}$ & 0.30 & $4.0\times 10^{-4}$ \\
& Screen 9 & $1.0\times 10^{-6}$ & 0.80 & $6.0\times 10^{-4}$ \\
\hline
\end{tabular}
\caption{Hyperparameter values used for each dataset.}
\label{table:hyperparameters}
\end{table}

Here we present the values picked for the hyperparameter $t$ for both the QLP-Even and QLP-Odd components, independently, and also the value of $\alpha$ used for the LO method \cite{pech2019link} for each network. We omitted $t$ for QLP-Even in P2P-Gnutella given that this is a bipartite network, and thus predictions based on even length paths have no meaning. The values were chosen by removing 10\% of the links from the training set in the first iteration of the 10-fold cross-validation procedure and maximizing the prediction precision for those removed links.

One immediate observation is that the values of $t$ for the predictions from QLP-Even and QLP-Odd are different by many orders of magnitude. From Eq. 3 in the main text we can write the prediction matrices for both components as follows
\begin{align}
    P^\text{even}&=|I-\frac{t^2}{2!}A^2+\frac{t^4}{4!}A^4-\frac{t^6}{6!}A^6+...|_{(ij)}^2\\
    P^\text{odd}&=|-tA+\frac{t^3}{3!}A^3-\frac{t^5}{5!}A^5+\frac{t^7}{7!}A^7-...|_{(ij)}^2,
\end{align}
where $|.|_{(ij)}^2$ denotes the entry-wise absolute value squared. In both cases the first term does not contribute to the predictions. The small values of $t$ in QLP-Even indicate that these predictions are best represented by the $A^2$ component in the series, while thes values observed for QLP-Odd indicate that these predictions typically benefit from the contributions of the higher-order powers of $A$.

\clearpage
\begin{figure}[t]
\centering
\includegraphics[width=\textwidth]{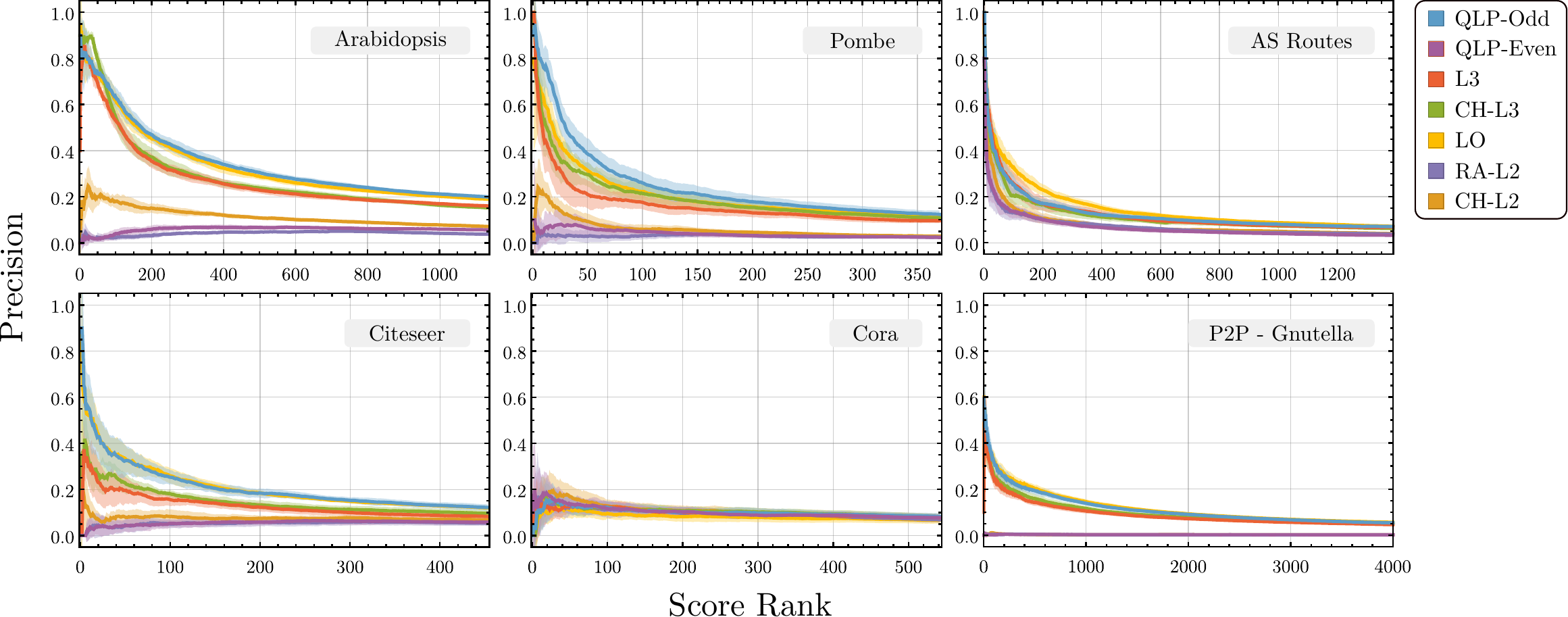}
\caption{\textbf{Extra cross-validation results: Precision.}}
\label{fig:5}
\end{figure}

Cumulative precision over the top ranked predictions (the top $0.05*Nk_\text{av}$ scores) for each network, averaged over a 10-fold cross validation procedure. The shaded regions correspond to the standard deviation. In each trial 10\% of the links were randomly removed and the remainder used as input to the link prediction methods. The networks used correspond to the PPI network Arabidopsis \cite{arabidopsis}, Pombe \cite{biogrid}, AS Routes \cite{asroutesp2p}, Citeseer \cite{citeseercora}, Cora \cite{citeseercora} and P2P-Gnutella \cite{asroutesp2p}. We note that P2P-Gnutella is a bipartite network, and thus the TCP based methods QLP-Even, RA-L2 and CH-L2 have null precision. The dataset parameters characterizing each network are shown in Table \ref{table:datasetparameters}, and the values selected for the optimal parameters $t$ in the QLP method and $\alpha$ in the LO method are shown in Table \ref{table:hyperparameters}.

\begin{table}[h!]
\centering
\begin{tabular}{cc|cc|ccccc}
& Dataset  & QLP-Even & QLP-Odd & LO & L3 & CH-L3 & RA-L2 & CH-L2  \\\hline
\multirow{5}{0.1cm}{\rotatebox{90}{\textbf{AUC-ROC}}}
& Arabidopsis & 0.661 & \textbf{0.776} & 0.792 & \textbf{0.808} & \textbf{0.808} & 0.574 & 0.574 \\
& Pombe & 0.670 & \textbf{0.685} & 0.727 & \textbf{0.729} & \textbf{0.729} & 0.563 & 0.563 \\
& AS Routes & 0.701 & \textbf{0.725} & \textbf{0.810} & 0.743 & 0.742 & 0.604 & 0.604 \\
& Citeseer & 0.638 & \textbf{0.675} & 0.703 & \textbf{0.743} & \textbf{0.743} & 0.672 & 0.672 \\
& Cora & 0.706 & \textbf{0.781} & 0.679 & \textbf{0.766} & \textbf{0.766} & 0.709 & 0.709 \\\hline
\multirow{5}{0.1cm}{\rotatebox{90}{\textbf{AUC-PR}}}
& Arabidopsis & 0.007 & \textbf{0.127} & \textbf{0.119} & 0.006 & 0.015 & 0.092 & 0.097 \\
& Pombe & 0.004 & \textbf{0.070} & \textbf{0.052} & 0.040 & 0.047 & 0.004 & 0.006 \\
& AS Routes & 0.007 & \textbf{0.028} & \textbf{0.029} & 0.026 & 0.023 & 0.008 & 0.011 \\
& Citeseer & 0.014 & \textbf{0.049} & \textbf{0.046} & 0.029 & 0.032 & 0.014 & 0.017 \\
& Cora & 0.023 & \textbf{0.024} & 0.017 & 0.022 & 0.023 & 0.022 & \textbf{0.024} \\\hline
\end{tabular}
\caption{\textbf{Extra cross-validation results: AUC.}}
\label{tab:extraauc}
\end{table}

Area under Curve (AUC) performance metrics for the datasets in Figure \ref{fig:5}, for both the receiver-operating curve (AUC-ROC) and precision-recall curve (AUC-PR). P2P-Gnutella is omitted due to a lack of computational resources to evaluate these metrics for the full network. The metrics were computed over the full set of potential predictions for each dataset in a 10-fold cross-validation procedure. Each value corresponds to the mean AUC over the ten iterations. The parameter values of $t$ and $\alpha$ for the QLP and LO methods, respectively, were the same shown in Table \ref{table:hyperparameters}. In bold are the best values for both QLP and the classical methods tested.

\clearpage
\begin{figure}[t]
\centering
\includegraphics[width=\textwidth]{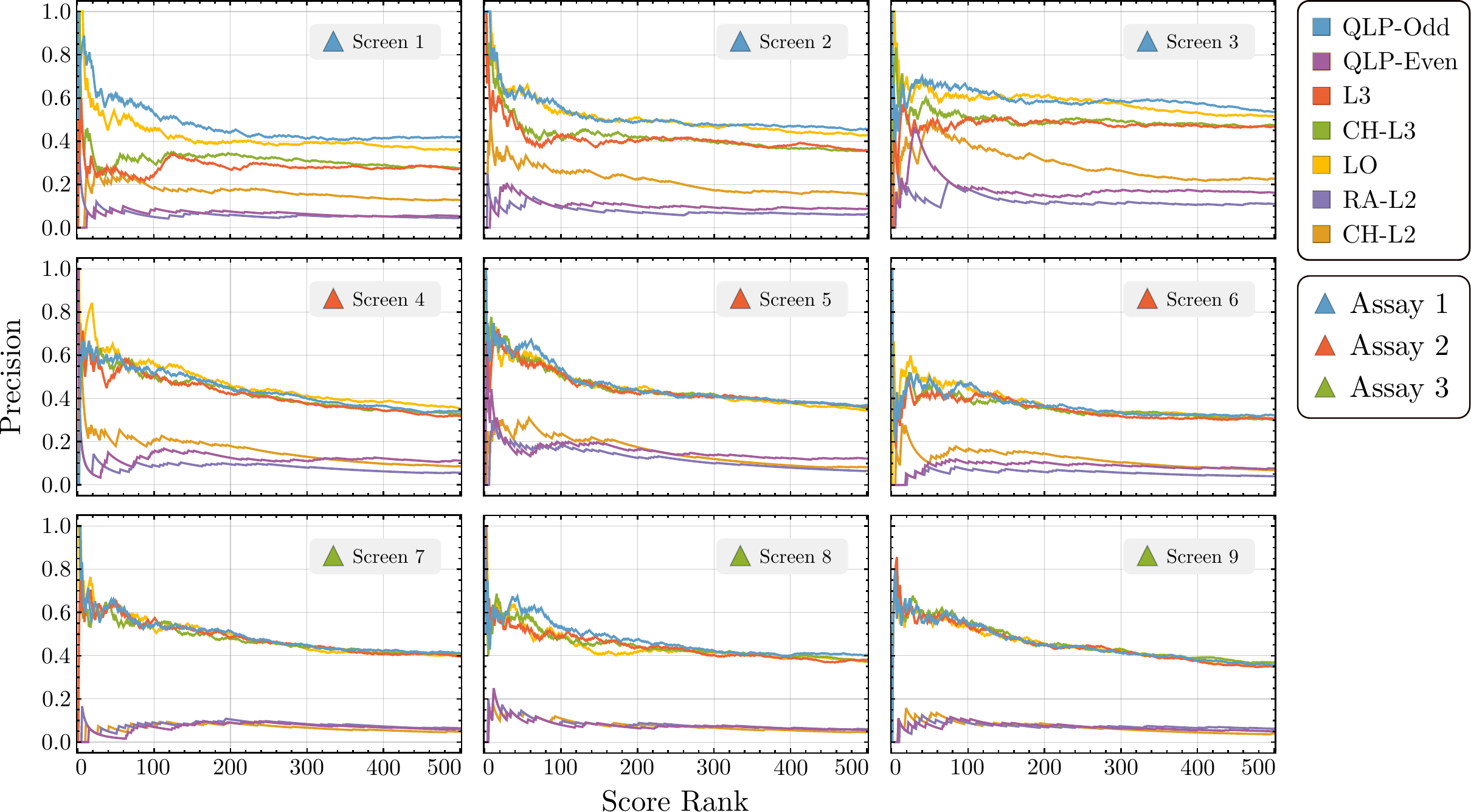}
\caption{\textbf{Prediction of missing PPIs in the human interactome validated against experimental screens.}}
\label{fig:6}
\end{figure}

The results presented correspond to the cumulative precision over the top 500 ranked predictions. The dataset used consists of nine different screens over a search space of human binary PPIs using a panel of three different assay versions \cite{luck2020interactome}, the most recent experimental study of the human interactome network, named HI-III-20. In this study the authors presented a reference interactome map of human binary protein interactions with 52,569 protein-protein interactions involving 8,275 proteins. This map was generated by screening a search space of roughly $90\%$ of the protein-coding genome a total of nine times with a panel of three different but complementary assay versions. For each of the nine plots we used the results of the respective screen as the input network to the link prediction methods and compared the predictions obtained to the PPIs identified in the remaining two screens from the same assay. For example, for the case of Screen 1, the predictions were compared with the PPIs identified in Screen 2 and Screen 3 combined. For the methods with a free parameter (QLP and LO) we randomly removed $50\%$ of the input dataset and optimized the method to best predict the removed links by maximizing the area under the precision curve over the top 500 score ranks. The optimized parameter was then used for the results displayed.

\end{document}